\documentclass[12pt]{iopart}
\expandafter\let\csname equation*\endcsname\relax
\expandafter\let\csname endequation*\endcsname\relax
\usepackage{amssymb}
\usepackage{amsmath}
\usepackage{amsthm}
\usepackage{amsfonts} 
\usepackage{bm}
\usepackage{physics}
\usepackage{graphics}
\usepackage{braket} 
\usepackage{graphicx}
\usepackage{dcolumn}

\begin{document}

\title[]{Grand canonical partition function of a serial metallic island system}

\author{Pipat Harata and Prathan Srivilai*}

\address{Theoretical Condensed Matter Physics Research Unit, Department of Physics,
Faculty of Science, Mahasarakham University, Khamriang Sub-District, Kantarawichai District, Mahasarakham 44150, Thailand}
\ead{prathan.s@msu.ac.th}
\vspace{10pt}
\begin{indented}
\item[]October 2021
\end{indented}

\begin{abstract}
We present a calculation of the grand canonical partition function of a serial metallic island system by the imaginary-time path integral formalism. To this purpose, all electronic excitations in the lead and island electrodes are described using Grassmann numbers. Coulomb charging energy of the system is represented in terms of phase fields conjugate to the island charges. By the large channel approximation, the tunneling action phase dependence can also be determined explicitly. Therefore, we represent the partition function as a path integral over phase fields with a path probability given in an analytically known effective action functional.  Using the result, we also propose a calculation of the average electron number of the serial island system in terms of the expectation value of winding numbers. Finally, as an example, we describe the Coulomb blockade effect in the two-island system by the average electron number and propose a method to construct the quantum stability diagram.
\end{abstract}

%
\vspace{2pc}

\section{Introduction}
\label{sec:intro}

Single-electron tunneling devices are standard tools in nano-science\cite{GrabertDevoret, Devoret1992}. They continue to attract attention because of their nanoscopic scale, low power dissipation, and new functionalities \cite{Likharev1999}. A single-electron tunneling device allows us to control the tunneling current at the single electron level. Electron transport is strongly affected by the charging effect at temperatures in the sub-Kelvin range \cite{Dittrich, Alhassid2000}. It is, therefore, crucial to include the charging energy into the tunneling processes in the single-electron tunneling devices \cite{Kastner}. The most widely studied device is the single-electron transistor (SET), consisting of a single island, two tunnel junctions, and a gate electrode controlling the electrostatic potential of the island. In theoretical studies, the imaginary-time path integral formalism is a powerful tool for describing the experimental data of the SETs with high accuracy \cite{Joyez1997, Goppert2000, walli02}. The findings on the SETs indicate that a close match between the theoretical model and its experimental realization exists for the single-electron transistor. In addition, one can apply the path integral Monte Carlo simulation \cite {Herrero1999, Werner2005, Lukyanov2006, Ceperley1995} to study the systems as the SET without the restriction of coupling parameters and temperature as perturbation theory  \cite{Grabert1994, Goppert2001, Altland2006} and semiclassical approximation\cite{Goppert2000, Grabert2000}, respectively. However, since the complexity of the single-electron tunneling devices depends on the number of the island, it is still unclear whether the theoretical calculation would agree with more complex experiments than that of the SET, i.e., the systems consist of many islands \cite {limbach05, Gaudreau2009, Granger2010}.

Due to quantum statistical properties that can be obtained from the system's partition function, this paper aims to calculate the grand canonical partition function of a finite serial metallic island system by the imaginary-time path integral approach \cite{Negele1987, Ambegaokar1982, Grabert1994}. The paper is organized as follows: Section 2 introduces the Hamiltonian of the serial metallic island system and some basic notations. This section deals with the path integral representation of the grand canonical partition function of the serial island system. To this purpose, we describe all electronic excitation in the lead and island electrodes employing Grassmann numbers and Coulomb charging of the island electrodes in terms of phase fields conjugate to the island charges. The general form of the path integrals over the phase and Grassmann fields is discussed. In Section 3, for further application, we propose a suitable form of the average electron number of the island system for the quantum Monte Carlo simulation and apply it to describe the Coulomb blockade effect in the two-island system as an example. Finally, we conclude and discuss possible extensions in Section 4.

\section{Model and the imaginary-time path integral}
\label{sec2}
 In this section, the grand canonical partition function and the effective action of a finite serial metallic island are derived. Section \ref{sec:model}, we first introduce the Hamiltonian model used to describe a serial metallic island system. In Section \ref{subsec:ansatz} deals with the imaginary-time path integral to represent the partition function of the system. The Coulomb action is then evaluated in Section \ref{subsec:Path over phase}. Path integral over Grassmann fields and the integration are discussed in Section \ref{subsec:Path over Grassmann} and Section \ref{subsec:integration}, respectively. Finally, in Section \ref{subsec:theSt}, the tunneling action is evaluated by the large channel approximation \cite{Grabert1994,Goppert1999}.  
 
\subsection{Hamiltonian model}
\label{sec:model} 

\begin{figure}
\centering
\includegraphics[width=0.75\textwidth]{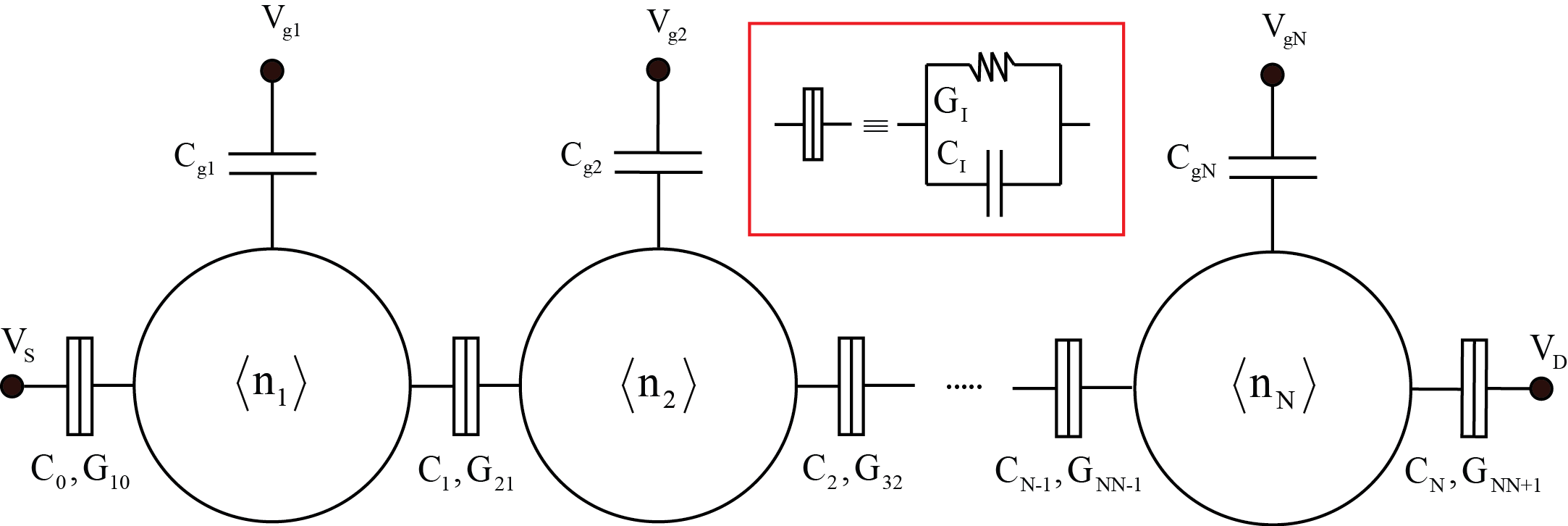}
\vspace{0.2cm}
\caption{This arrangement is biased by the voltage difference $V_S-V_D$, where $V_S$ and $V_D$ are the source and drain voltage, respectively. The external gate voltage can tune the electrostatic potential on the island $I$, i.e., $V_{gI}$, which couples directly to the island by the capacitance $C_{gI}$. The average excess electron number on the island $I$ is denoted by $\left\langle {{n_I}} \right\rangle$. All tunneling junctions are represented by a conductor connected in parallel with a capacitor.} 
\label{fig:fig_model}
\end{figure}

A circuit diagram of the finite serial island system is depicted in Fig.\ref{fig:fig_model}. Microscopically, this system can be theoretically modeled by the Hamiltonian \cite{GrabertDevoret}
\begin{eqnarray} \label{Hamiltonian}
H=H_{B}+H_{T} + H_{C}\, ,
\end{eqnarray}
where the term $H_{B}$ describes the conduction electrons of the leads and the islands
\begin{eqnarray}
\label{HsubB} H_{B}=\sum_{Jk\sigma}\epsilon_{Jk\sigma}c_{Jk\sigma}^{\dag}c_{Jk\sigma}^{}+
\sum_{Ik\sigma}\epsilon_{Ik\sigma}d_{Ik\sigma}^{\dag}d_{Ik\sigma}^{}\, ,
\end{eqnarray}
respectively. Here $\epsilon_{Jk\sigma}$ is the energy of an electron with longitudinal wave vector $k$ in channel $\sigma$ of lead $J$, where $J \in \{S,D\}$. The channel index $\sigma$ includes the transversal and spin quantum numbers. 
$c_{Jk\sigma}$ and $c_{Jk\sigma}^{\dag}$ are the corresponding electron annihilation and creation operators, respectively. Likewise, $\epsilon_{Ik\sigma}$ denotes the energy of an
electron with the longitudinal wave vector $k$ in channel $\sigma$ of island $I$, where $I \in \{1,2,...,N\}$, and $d_{Ik\sigma}$ and $d_{Ik\sigma}^{\dag}$ are the associated electron annihilation and creation operators, respectively.

The second term in Eq.(\ref{Hamiltonian}) describes the tunneling of electrons across the finite tunneling junctions of the circuit as shown in Fig.\ref{fig:fig_model},
\begin{eqnarray}
\label{Htunneling}
H_{T}=\sum_{kq\sigma}&\Big[&d_{1k\sigma}^{\dag}t_{1Skq \sigma}{\rm e}^{-i\varphi_{1}}c_{Sq\sigma}^{}
 +\sum_{I=2}^{N}d_{Ik\sigma}^{\dag}t_{II-1kq \sigma}{\rm e}^{-i(\varphi_{I}-\varphi_{I-1})}c_{I-1q\sigma}^{}\\ 
 &&+c_{Dk\sigma}^{\dag}t_{DNkq \sigma}{\rm e}^{i\varphi_{N}}d_{Nq\sigma}^{}+\hbox{H.c.}\label{HsubT}\Big] \nonumber,
\end{eqnarray}
where $t_{1Skq\sigma}$, $t_{II-1kq\sigma}$, and $t_{DNkq \sigma}$ denote electron tunneling amplitude of the tunneling junctions labeled by $1S, II-1$, and $DN$, respectively. For example, $t_{1Skq\sigma}$ is the amplitude for an electron in state $|q\sigma\rangle$ on lead $S$ to tunnel onto island $1$ with final state $|k\sigma\rangle$. In addition, we have introduced the phase operators $\varphi_{I}$  conjugate to the number operators $n_{I}$ of excess charges on the islands $I$, i.e. $[n_{I},\varphi_{I}]=i$. Accordingly, the charge shift operator ${\rm e}^{-i\varphi_{I}}$ add one charge to the island $I$. Here and in the following, we have defined  $\varphi_S=\varphi_D=0$ since $V_S=V_D=0$.

The last term in Eq.(\ref{Hamiltonian}) is Coulomb charging energy expressed as\cite{GrabertDevoret, Fujisawa2002}
\begin{eqnarray}\label{Ec0}
   H_{C}=\sum\limits_{I=1}^{N}{{E_{II}}{( {n_I}-{n_{0I}})}^{2}}+2\sum\limits_{I<I'}^{N}{{E_{II'}}( {n_I}-{n_{0I}})( {n_{I'}}-{n_{0I'}})},
\end{eqnarray}
where $n_I$ is the (excess) electron number in the island $I$. For the island $I$ the continuous charge induced by the gate voltage $I$ is denoted by $n_{0I}$ expressed by $n_{0I}=C_{gI}V_{gI}/e$. The variable $n_{0I}$ is non-integer and just a parameter of the external voltage. The coefficients $E_{II}$ and $E_{II'}$ are the matrix elements of the matrix $\mathbf{E}_{CN}$ defined by
\begin{eqnarray}\label{ECN}
  \mathbf{E}_{CN}
=
  \frac{{{e}^{2}}}{2}\mathbf{C}_{N}^{-1},
\end{eqnarray}
where $\mathbf{C}_N^{-1}$ is the inverse matrix of $\mathbf{C}_{N}$
\begin{eqnarray} \label{Cn}
 \mathbf{C}_{N}
=
\begin{pmatrix}
  C_{11} & C_{12} & \cdots & C_{1N} \\
  C_{21} & C_{22} & \cdots & C_{2N} \\
  \vdots & \vdots & \vdots & \vdots \\
  C_{N1} & \cdots & \cdots & C_{NN}
\end{pmatrix},
\end{eqnarray}
and the matrix elements of $C_{II'}$ are defined by the conditions:
\begin{eqnarray}\label{element matrix C}
    C_{II'}=
    \begin{cases}
      0, & \text{if} \,\,\,\,\,|I-I'|>1 \\
      C_{\sum\nolimits I}, & \text{if} \,\,\,\,\, I=I' \\
      -C_{I'}, & \text{if} \,\,\,\,\,|I-I'|=1,
    \end{cases}
\end{eqnarray}
where $C_{\sum\nolimits I}$ is the sum of all capacitors connected with the island $I$, i.e., $C_{\sum I}=C_I+C_{I-1}+C_{gI}$. According to the conditions in Eq.(\ref{element matrix C}), one obtains $C_{II'}=C_{I'I}$.

\subsection{Path integral representation of the partition function}
\label{subsec:ansatz}
The goal of this section is a path integral representation of the grand partition function
\begin{eqnarray}
\label{partitionf}
Z={\rm tr}\, \{{\rm e}^{-\beta (\hat{H}-\mu\hat{N})}\}
\end{eqnarray}
of the island system suitable for the evaluation, where $\mu$ denotes the chemical potential of the system and $\hat{N}$ is the particle number operator. The Hilbert space of the Hamiltonian in Eqs.(\ref{Hamiltonian})--(\ref{Ec0}) is the product of the space spanned by the state $\ket{n}$, or equivalently the phase state $\ket{\varphi}$, and the Fock space of the quasi--particles. In order to trace over the quasi--particle excitation, we introduce Grassmann numbers $\zeta_{Jk\sigma}$ and $\theta_{Ik\sigma}$ corresponding to the electron annihilators $c_{Jk\sigma}$ and $d_{Ik\sigma}$, respectively. For notation convenience, the Grassmann numbers for electronic lead states are combined to a single vector $\vec{\zeta}=(\zeta_{Sk\sigma},\zeta_{Dq\sigma})^{T}$. Likewise, $\vec{\theta}=(\theta_{1q\sigma},\theta_{2q\sigma},\ldots,\theta_{Nq\sigma})^{T}$ combines all Grassmann numbers for electronic island states. To trace over the island charges, we work in the phase representation and introduce the vector $\vec{\varphi}=(\varphi_1,\ldots,\varphi_N)^{T}$. 
The partition function in Eq.(\ref{partitionf}) may then be written as
\begin{eqnarray} \label{partitionfu}
Z=\int D\mu(\vec{\varphi}) \int D\mu(\vec{\zeta})\int D\mu(\vec{\theta}) {\rm e}^{-\vec{\zeta}^{\ast}\cdot\vec{\zeta}-\vec{\theta}^{\ast}\cdot\vec{\theta}}\langle
\vec{\varphi},-\vec{\zeta},-\vec{\theta}|\, {\rm e}^{-\beta H}\,| \vec{\varphi},\vec{\zeta},\vec{\theta}\rangle \,,
\end{eqnarray}
where we have dropped the term $(\mu\hat{N})$ in the partition function. This term is further discussed this term in Section \ref{subsec:Path over Grassmann}. Here $\vert\vec{\varphi}\rangle$ stands for the product of phase states $\vert\varphi_I\rangle$. The shorthand notations $\vert\vec{\zeta}\rangle$ and $\vert\vec{\theta}\rangle$ stand for the product of fermion coherent states $\vert\zeta_{Jk\sigma}\rangle$ and $\vert\theta_{Ik\sigma}\rangle$, respectively. The shorthand integration
\begin{eqnarray}
    \int D\mu(\vec{\zeta})\equiv  \int\prod_{Jk\sigma}d\zeta_{Jk\sigma}^{\ast}\, d\zeta_{Jk\sigma}^{} ,\label{intgamma}
\end{eqnarray}
and 
\begin{eqnarray}
    \int D\mu(\vec{\theta})\equiv  \int\prod_{Iq\sigma}d\theta_{Iq\sigma}^{\ast}\, d\theta_{Iq\sigma}^{} \label{intdelta}
\end{eqnarray}
are over all Grassmann numbers related with the lead and island states, respectively. The short hand integration
\begin{eqnarray}\label{intphi}
\int D\mu(\vec{\varphi})\, \equiv\, \left(\frac{1}{2\pi}\right)^2\int\prod_{I}\, d\varphi_I
\end{eqnarray}
integrates over the phases $\varphi_I$ that are $2\pi$-periodic. Further, we have introduced the notations
\begin{eqnarray}\label{GammaGamma}
    \vec{\zeta}^{\ast}\cdot\vec{\zeta} =
\sum_{Jk\sigma}\zeta^{\ast}_{Jk\sigma}\zeta_{Jk\sigma}^{} ,
\end{eqnarray}
and 
\begin{eqnarray}
    \vec{\theta}^{\ast}\cdot\vec{\theta} =\sum_{Iq\sigma}\theta^{\ast}_{Iq\sigma}\theta_{Iq\sigma}^{} \, .\label{DeltaDelta}
\end{eqnarray}
The derivation of the path integral expression can be done as usual by multiple insertions of the closure relation \cite{Negele1987}
\begin{eqnarray}
\label{closure}
1 = \int D\mu(\vec{\varphi}) \int D\mu(\vec{\zeta})\int D\mu(\vec{\theta}){\rm e}^{-\vec{\zeta}^{\ast}\cdot\vec{\zeta}-\vec{\theta}^{\ast}\cdot\vec{\theta}}\, \ket{\vec{\varphi},\vec{\zeta},\vec{\theta}} \bra{\vec{\varphi},\vec{\zeta},\vec{\theta} },
\end{eqnarray}
in the product space. The imaginary--time step is introduced as $\Delta = \beta/P$, where we have used $\hbar=1$ through this paper, and $P$ denotes the Trotter number. For each imaginary time segment $\Delta_{j}=\tau_{j}-\tau_{j-1},$ one can calculate the short-time propagator
\begin{eqnarray} \label{shorttimepropagator}
\langle\vec{\varphi}_{j},\vec{\zeta}_{j},\vec{\theta}_{j}|\, {\rm e}^{-\Delta_{j}H}\,
|\vec{\varphi}_{j-1},\vec{\zeta}_{j-1},\vec{\theta}_{j-1}\rangle &=&\langle\vec{\zeta}_{j},\vec{\theta}_{j}|\, {\rm
e}^{-\Delta_{j}\left[H_B+H_T(\vec{\varphi}_{j-1})\right]}\, |\vec{\zeta}_{j-1},\vec{\theta}_{j-1}\rangle\nonumber \\ &&\times\langle\vec{\varphi}_{j}|{\rm e}^{-\Delta_{j}H_{C}}|\vec{\varphi}_{j-1}\rangle,
\end{eqnarray}
where $\vec{\varphi}_{j}=(\varphi_{1}(\tau_{j}),\varphi_{2}(\tau_{j}),\ldots,\varphi_{N}(\tau_{j}))^{T}$  and $\vec{\varphi}_{j-1}=(\varphi_{1}(\tau_{j-1}),\varphi_{2}(\tau_{j-1}),\ldots,\varphi_{N}(\tau_{j-1}))^{T}$ and  $\vec{\zeta}_{j}$ and $\vec{\theta}_{j}$ are the Grassmann variables at time $\tau_{j}$. Further, $H_T(\vec{\varphi}_j)$ refers to the Hamiltonian in Eq.(\ref{Htunneling}) with the phase operators $\varphi_I$ replaced by  $\varphi_I(\tau_j)$. Making use of the form in Eq.(\ref{shorttimepropagator}) of the short time propagator in the Trotter break-up for the partition function, we obtain
\begin{eqnarray}\label{pf1}
Z &=&\int\limits_{0}^{2\pi} \prod_{I=1}^N \Big[ d\varphi_{I,P}\ldots
d\varphi_{I,0}\, \delta(\varphi_{I,P}-\varphi_{I,0}) \Big] \\ 
&&\times \langle\vec{\varphi}_{P}|\,{\rm
e}^{-\Delta_{P}H_{C}}\,|\vec{\varphi}_{P-1}\rangle\ldots \langle\vec{\varphi}_{1}|\,{\rm e}^{-\Delta_{1}H_{C}}\,|\vec{\varphi}_{0}\rangle \,Z_{BT}[\vec{\varphi}] \nonumber,
\end{eqnarray}
where $Z_{BT}[\vec{\varphi}]$ is the trace over the Grassmann fields discussed in more detail in Section \ref{subsec:Path over Grassmann}.
\subsection{Path integral over phase fields and Coulomb action}
\label{subsec:Path over phase}
The representation of the partition function in Eq.(\ref{pf1}) can be evaluated by inserting the closure relation
\begin{eqnarray}\label{closure1}
1 = \int\limits_0^{2\pi } {d\varphi \left| \varphi  \right\rangle } \left\langle \varphi  \right| = \sum\limits_{n = - \infty }^\infty  {\left| n \right\rangle } \left\langle n \right|,
\end{eqnarray}
where $n$ is an integer number into the short time propagator in the last term in Eq.(\ref{shorttimepropagator}). We have
\begin{eqnarray}
\langle\vec{\varphi}_{j}|{\rm e}^{-\Delta_{j}H_{C}}|\vec{\varphi}_{j-1}\rangle
&=&\sum_{\vec{n}}\langle\varphi_{1,j}|n_{1}\rangle\, \ldots\langle\varphi_{N,j}|n_{N}\rangle \\ \nonumber \\ \nonumber
&&\times{\rm e}^{-
\Delta_{j}H_{C}(n_{1},\ldots,n_{N})} \langle n_{1}|\varphi_{1,j-1}\rangle\ldots\langle
n_{N}|\varphi_{N,j-1}\rangle\nonumber \\ \nonumber
&=&\frac{1}{({2\pi})^{N}}\sum_{\vec{n}}{\rm e}^{-\Delta_{j}H_{C}(n_{1},\ldots,n_{N})}{\rm e}^{-i\sum_{I}n_{I}(\varphi_{I,j}-\varphi_{I,j-1})} \nonumber,
\end{eqnarray}
where we have defined
\begin{eqnarray}
    \sum_{\vec{n}} \equiv \sum_{n_1=-\infty}^\infty...\sum_{n_N=-\infty}^\infty.
\end{eqnarray}
$H_C(n_{1},\ldots,n_{N})$ is the Coulomb Hamiltonian in Eq.(\ref{Ec0}) for given eigenvalues of the electron number operators $n_I$ and $
\left\langle {{{n_I}}}
 \mathrel{\left | {\vphantom {{{n_I}} {{\varphi _{I,j}}}}}
 \right. \kern-\nulldelimiterspace}
 {{{\varphi _{I,j}}}} \right\rangle  = {\left( {2\pi } \right)^{ - 1/2}}\exp \left( {i{n_I}{\varphi _{I,j}}} \right)$. By means of the Poisson re-summation formula \cite{Negele1987},
\begin{eqnarray}\label{Poisson} \sum_{n=-\infty}^{\infty} f(n)=\sum_{k=-\infty}^{\infty} \ \int\limits_{-\infty}^{\infty}\!\! d
n \, {\rm e}^{-2\pi i k n}f(n)\, ,
\end{eqnarray}
and the matrix in Eq.(\ref{ECN}) corresponding with $H_C$, the short time propagator can be transformed to read
\begin{eqnarray}\label{Poisson1}
\langle\vec{\varphi}_{j}|{\rm e}^{-\Delta_{j}H_{C}}|\vec{\varphi}_{j-1}\rangle 
&=& \frac{1}{(2\pi)^{N}}\sum_{\vec{k}_{j}}\int\limits_{-\infty}^{\infty}
d\tilde{n}_{1}\ldots\int\limits_{-\infty}^{\infty} d\tilde{n}_{N}\\
&&\times\exp\left[{-i\Delta _j\sum_{I=1}^N\tilde{n}_{I}\Delta\varphi_{I}}-{\Delta_{j}\vec{\tilde n}_I^T
 \mathbf{E}_{CN}
\vec{\tilde n}_I}\right]\, \nonumber,
\end{eqnarray}
where we have defined
\begin{eqnarray}
    \sum_{\vec{k}_{j}} \equiv \sum_{k_{1,j}=-\infty}^\infty...\sum_{k_{N,j}=-\infty}^\infty,
\end{eqnarray}
$\vec{\tilde n}_I = {\left( {{{\tilde n}_1},...,{{\tilde n}_N}} \right)^T}$ with $\tilde{n}_{I}\equiv (n_{I}-n_{I0}) $, and $\Delta {\varphi _I} = \left( {{\varphi _{I,j}} - {\varphi _{I,j - 1}} + 2\pi {k_{I,j}}} \right)/{\Delta _j}$. Evaluating the Gaussian integrals for matrix form \cite{Negele1987}
\begin{eqnarray}\label{Gaussian}
\int\limits_{ - \infty }^\infty  {d{n_1}...d{n_N}} \exp[ { - \lambda( {{\vec n^T}{\mathbf E_{CN}} \vec n + {\vec n^T}\vec J})}] &=& {( {\frac{\pi }{\lambda }} )^{N/2}}[ {\det {\mathbf E_{CN}}]^{ - \frac{1}{2}}} \\
&&\times\exp[ {({\frac{\lambda}{4}}){\vec J}^T{\mathbf E_{CN}^{-1}}\vec J}] \nonumber,
\end{eqnarray}
where $\vec{n} = {\left( {{{\tilde n}_1},...,{{\tilde n}_N}} \right)^T}$, $\vec J = {\left( {\Delta {\varphi _1},...,\Delta {\varphi _N}} \right)^T}$, and $\lambda$ is a constant,
one obtains
\begin{eqnarray}\label{Poisson2}
\langle\vec{\varphi}_{j}|{\rm e}^{-\Delta_{j}H_{C}}|\vec{\varphi}_{j-1}\rangle
= {\mathcal{N}_j} \sum_{\vec{k}_{j}} {\exp \left[ {\frac{{ - {\Delta _j}}}{4}\Delta \vec\varphi _j^T\mathbf {E}_{CN}^{-1}\Delta {\vec\varphi _j} - i{\Delta _j}\vec {\bm n}_g^T\Delta {\vec\varphi _j}} \right]}
\end{eqnarray}
where $\mathcal{N}_{j}$ is the normalization constant for the time segment $ j $; 
\begin{eqnarray}
{\mathcal{N}_j} = \frac{1}{{{{\left( {2\pi } \right)}^N}}}{\left( {\frac{\pi }{{{\Delta _j}}}} \right)^{N/2}}{\left[ {\det {\mathbf {E}_{CN}}} \right]^{ - \frac{1}{2}}},
\end{eqnarray}
that can be incorporated into the path integral measure.
We now use the freedom to relabel the summations over the winding numbers $k_{I}$ and to transform the integrals over
$\varphi_{I}$. Instead of summations over $ k_{I,j},\ldots,k_{I,P} $, we sum over
\begin{eqnarray}\label{kprime}
k^{\prime}_{I,n}=\sum_{j=1}^{n} k_{I,j}\quad\mbox{for}\quad n=1,\ldots,P\, ,
\end{eqnarray}
with the consequence that
\begin{eqnarray}\label{DeltaPhi2}
 \Delta\varphi_{I,j}=\frac{\varphi_{I,j}+2\pi k^{\prime}_{I,j}-\varphi_{I,j-1}-2\pi k^{\prime}_{I,j-1}}{\Delta_{j}}.
\end{eqnarray}
Using $\varphi^{\prime }_{I,j}=\varphi_{I,j}+2\pi
k^{\prime}_{I,j}$ and dropping the primes as a convenient integration variable, we can rewrite the partition function in Eq.(\ref{pf1}) as
\begin{eqnarray}\label{pf2}
 Z&=&\mathcal{N}\prod\limits_{I=1}^N\Bigg[\sum_{k_{I,j}=-\infty}^{\infty}
\int\limits_{2\pi k_{I,P}}^{2\pi(k_{I,P}+1)}d\varphi_{I,P}\ldots\int\limits_{2\pi
k_{I,1}}^{2\pi(k_{I,1}+1)}d\varphi_{I,1} \int\limits_{0}^{2\pi}d\varphi_{I,0} \nonumber
\\
&&\times\delta(\varphi_{I,P}-\varphi_{I,0}-2\pi k_{I,P})\Bigg] \\ &&\times\exp[\sum_{j=1}^{P}\Delta_{j}(\frac{1}{4}\Delta\vec{\varphi}^{T}_{j}\mathbf {E}_{CN}^{-1}\Delta\vec{\varphi}_{j}+i\vec{ n}^{T}_{g}\cdot\Delta\vec{\varphi}_{j})] Z_{BT}[\vec{\varphi}]\nonumber,
\end{eqnarray}
where we have introduced the vectors
\begin{eqnarray}\label{vecphi}
    \Delta\vec{\varphi}_{j}=\left(\frac{\varphi_{1,j}-\varphi_{1,j-1}}{\Delta_{j}},\frac{\varphi_{2,j}-\varphi_{2,j-1}}{\Delta_{j}},...,
\frac{\varphi_{N,j}-\varphi_{N,j-1}}{\Delta_{j}}\right)^{T},
\end{eqnarray}
and $\vec{n}_{g}=\left(n_{01},n_{02},... n_{0N}\right)^{T}$. The normalization factor $\mathcal{N}\equiv\prod_{j=1}^{P-1} \mathcal{N}_{j}$.

With the exception of $k_{P}$, the sums over $k_{j}$ can be incorporated in the integrals over $\varphi_{I,j}$. This simplifies
the expression in Eq.(\ref{pf2}) further, and we get in the continuum limit, i.e., ($P\rightarrow\infty ,\ \Delta_{j}\rightarrow0$) which is a path integral over all imaginary-time paths $\vec{\varphi}(\tau)=(\varphi_1(\tau),\varphi_2(\tau),...,\varphi_N(\tau))^{T}$ in the time interval $(0,\beta)$ with arbitrary winding numbers
\begin{eqnarray}\label{pf3}
Z&=&\mathcal{N}\prod\limits_{I=1}^N\bigg[\sum_{k_{I}=-\infty}^{\infty}\int\limits_{\varphi_{I}(0)}^{\varphi_{I}(\beta)+2\pi
k_{I}}D[\varphi_{I}(\tau)]\bigg]{\rm e}^{-S_{C}[\vec{\varphi}(\tau)]}\, Z_{BT}[\vec{\varphi}]\,,
\end{eqnarray}
and the Coulomb action of the serial island system is expressed by
\begin{eqnarray}\label{Sc}
    S_{C} [\vec{\varphi}(\tau)]=\int_{0}^{\beta} d\tau
\left[\frac{1}{4}\dot{\vec{\varphi}}^{T}\mathbf {E}_{CN}^{-1}\,\dot{\vec{\varphi}}+i (\vec{ n}^{T}_{g}\cdot\dot{\vec{\varphi}})\right]\, ,
\end{eqnarray}
where
$\dot{\vec{\varphi}}(\tau_{j})=\left(\dot{\varphi}_1(\tau_{j}\right),\dot{\varphi}_2(\tau_{j}),...,\dot{\varphi}_N(\tau_{j}))^{T}$ is the continuum limit of $\Delta\vec{\varphi}_{j}$ for
$\Delta_{j}\rightarrow 0$ and $\mathbf {E}_{CN}^{-1}$ is the inverse matrix of the matrix $\mathbf{E}_{CN}$ in Eq.(\ref{ECN}).
\subsection{Path integral over Grassmann fields}
\label{subsec:Path over Grassmann}
 In this section, we have to evaluate the partition function in Eq.(\ref{pf3}) wherein $Z_{BT}$ involves the integration over the Grassmann numbers. In the Trotter break-up discussed in the previous section, the imaginary-time segment $\Delta_j$ gives rise to the short time propagator in Eq.(\ref{shorttimepropagator}) with the fermions short time propagator
\begin{eqnarray}
\langle\vec{\zeta}_{j},\vec{\theta}_{j}|\, {\rm e}^{-\Delta_{j}\left[H_B+H_T(\vec{\varphi}_{j-1})\right]}\,|\vec{\zeta}_{j-1},\vec{\theta}_{j-1}\rangle.
\end{eqnarray}
Within this short time propagator, the electron annihilation (creation) operators in the Hamiltonians $H_B$ in Eq.(\ref{HsubB}) and $H_T$ in Eq.(\ref{Htunneling}) can be replaced for small $\Delta_j$ by the corresponding (conjugate) Grassmann numbers. Accordingly, one finds
\begin{align}\label{stpFermion}
\langle\vec{\zeta}_{j},\vec{\theta}_{j}|\, {\rm e}^{-\Delta_{j}\left[H_B+H_T(\vec{\varphi}_{j-1})\right]}\,
|\vec{\zeta}_{j-1},\vec{\theta}_{j-1}\rangle
&=\langle\vec{\zeta}_{j},\vec{\theta}_{j}|\vec{\zeta}_{j-1},\vec{\theta}_{j-1}\rangle \\
&\times\exp\bigg\{-\Delta_{j}\Big[ H_{B,j,j-1}+ H_{T,j,j-1}(\vec{\varphi}_{j-1})\Big]\bigg\}\nonumber,
\end{align}
where
\begin{eqnarray}
H_{B,j,j-1}=\sum_{Jk\sigma}\epsilon_{Jk\sigma}\zeta^*_{Jk\sigma,j}\zeta_{Jk\sigma,j-1}
+ \sum_{Ik\sigma}\epsilon_{Ik\sigma}\theta^*_{Ik\sigma,j}\theta_{Ik\sigma,j-1}\,,
\end{eqnarray}
and
\begin{eqnarray}
H_{T,j,j-1}(\vec{\varphi}_{j-1})&=&\sum_{kq\sigma}\Big[\theta_{1k\sigma,j}^{*}t_{1Skq \sigma}\, {\rm e}^{-i\varphi_{1,j-1}}\zeta_{Sq\sigma,j-1} \\
&&+\sum_{I=2}^{N} \theta_{Ik\sigma,j}^{*}t_{II-1kq\sigma}\, {\rm e}^{-i(\varphi_{I,j-1}-\varphi_{I-1,j-1})}\theta_{I-1q\sigma,j-1}\nonumber\\
&&+\zeta_{Dk\sigma,j}^{*}t_{DNkq \sigma}\, {\rm e}^{i\varphi_{N,j-1}}\theta_{Nq\sigma,j-1}+ \hbox{H.c.}\Big]\nonumber\,.
\end{eqnarray}
Combining the exponential factor in Eq.(\ref{stpFermion}) with the factor $\exp\left(-\vec{\zeta}^{*}_{j}\cdot\vec{\zeta}_j-\vec{\theta}_{j}^{*}\cdot\vec{\theta}_{j}\right)$ stemming from the closure relation in Eq.(\ref{closure}), and the scalar product \newline $\langle\vec{\zeta}_{j},\vec{\theta}_{j}|\vec{\zeta}_{j-1},\vec{\theta}_{j-1}\rangle
=\exp\left(\vec{\zeta}_{j}^{*}.\vec{\zeta}_{j-1}+\vec{\theta}_{j}^{*}.\vec{\theta}_{j-1}\right)$, we obtain for each time segment $\Delta_j$ a factor
\begin{equation}
\exp\bigg\{-\Delta_{j}\bigg[ \vec{\zeta}_{j}^{*}\cdot\frac{\vec{\zeta}_j-\vec{\zeta}_{j-1}}{\Delta_j}+\vec{\theta}^{*}_{j}\cdot\frac{\vec{\theta}_j-\vec{\theta}_{j-1}}{\Delta_j}
+ H_{B,j,j-1}+ H_{T,j,j-1}(\vec{\varphi}_{j-1})\bigg]\bigg\}\, .
\end{equation}
In the continuum limit, these factors and the integration over the
Grassmann numbers $\vec{\zeta}_j$ and $\vec{\theta}_j$ are combined to the path integral
\begin{eqnarray}\label{ZTB}
Z_{BT}[\vec{\varphi}]=\int D\mu(\vec{\zeta})\int D\mu(\vec{\theta})\ {\rm e}^{-S_{BT}[\vec{\zeta},\vec{\theta},\vec{\varphi}]},
\end{eqnarray} where
\begin{eqnarray}\label{SBT}
S_{BT}[\vec{\zeta},\vec{\theta},\vec{\varphi}]=S_{\rm lead}[\vec{\zeta}]+S_{\rm isl}[\vec{\theta}]+S_{T}[\vec{\zeta},\vec{\theta},\vec{\varphi}],
\end{eqnarray}
with the lead action
\begin{eqnarray}\label{Slead}
S_{\rm lead}[\vec{\zeta}(\tau)]=\int_{0}^{\beta}d\tau\sum_{Jk\sigma}
\Big[\zeta_{Jk\sigma}^{\ast}(\tau)(\partial_{\tau}-\mu+\epsilon_{Jk\sigma})\zeta_{Jk\sigma}(\tau)\Big] \equiv \zeta^{\ast} g_J^{-1} \zeta\, ,
\end{eqnarray} 
and the island action
\begin{eqnarray}\label{Sisl}
S_{\rm isl}[\vec{\theta}(\tau)]=\int_{0}^{\beta}d\tau\sum_{Iq\sigma}
\Big[\theta_{Iq\sigma}^{\ast}(\tau)(\partial_{\tau}-\mu+\epsilon_{Iq\sigma})\theta_{Iq\sigma}(\tau)\Big] \equiv \theta^{\ast} G_I^{-1} \theta\, ,
\end{eqnarray}
where $\partial_{\tau}\equiv({\vec{x}_j-\vec{x}_{j-1}})/{\Delta_j}$ for any variable. We have defined the shorthand notations including the integration over imaginary-time in the vector multiplication. The inverse of the Green's functions of electrons in the lead $(J)$ and the island $(I)$ are denoted by $g_J^{-1}$ and $G_I^{-1}$, respectively. We will discuss the Green's functions in more detail in Section \ref{subsec:integration}.
The tunneling term reads
\begin{align}\label{ST}
S_{T}[\vec{\zeta},\vec{\theta},\vec{\varphi}]&=\int_{0}^{\beta}d\tau\Bigg[\sum_{kq\sigma}\theta_{1k\sigma}^{\ast}t_{1Skq \sigma}\, {\rm e}^{-i\varphi_{1}}\zeta_{Sq\sigma} +\sum_{I=2}^{N}\sum_{kq\sigma}\theta_{Ik\sigma}^{\ast}t_{II-1kq\sigma}\, {\rm e}^{-i(\varphi_{I}-\varphi_{I-1})}\theta_{I-1q\sigma}\nonumber\\
&+\sum_{kq\sigma} \zeta_{Dk\sigma}^{\ast}t_{DNkq \sigma}\, {\rm e}^{i\varphi_{N}}\theta_{Nq\sigma}+\hbox{H.c.}\Bigg].
\end{align}



Inserting the result in Eq.(\ref{ZTB}) into the representation of the partition function in Eq.(\ref{pf3}), we obtain a path integral representation of the partition function which will be the basis for the analysis in the section following.
We close this section with a summary of the path integral formulation for the partition function 
\begin{equation}\label{Zchi}
Z[\vec{\zeta},\vec{\theta},\vec{\varphi}]=\mathcal{N}\prod\limits_{I=1}^N\bigg[\sum_{k_{I}=-\infty}^{\infty}\int\limits_{\varphi_{I}(0)}^{\varphi_{I}(\beta)+2\pi
k_{I}}D[\varphi_{I}(\tau)]\bigg]\int D\mu(\vec{\zeta})\int D\mu(\vec{\theta})\ {\rm e}^{-S[\vec{\zeta},\vec{\theta},\vec{\varphi}]}\, ,
\end{equation}
where
\begin{equation}\label{Stotal}
S[\vec{\zeta},\vec{\theta},\vec{\varphi}]=S_C[\vec{\varphi}]+S_{BT}[\vec{\zeta},\vec{\theta},\vec{\varphi}]\, .
\end{equation}
Introducing vectors $\vec{\zeta}_J=(\ldots,\zeta_{Jk\sigma},\ldots)^{T}$, $\vec{\theta}_I=(\ldots,\theta_{Ik\sigma},\ldots)^{T}$ and
combining Eqs.(\ref{SBT})--(\ref{ST}), we found that
\begin{eqnarray} \label{Selectron}
S_{BT}[\vec{\zeta},\vec{\theta},\vec{\varphi}]
&=& S_{\rm lead}[\vec{\zeta}]+ S_{\rm isl}[\vec{\theta}]+\int_{0}^{\beta}d\tau
\Big[\theta_{1}^{*}\,\Lambda_{1S}^{}\zeta_{S}\\
&&+\sum_{I=2}^N(\theta_{I}^{*}\,\Lambda_{II-1}\theta_{I-1})+\zeta_{D}^{*}\, \Lambda_{DN}\theta_{N}^{}+\hbox{H.c.}\Big], \nonumber
\end{eqnarray}
where we have introduced the tunneling matrices
\begin{eqnarray}\label{Lamda def}
 {{\Lambda }_{1S}}&=&{{t}_{1Skq\sigma}}{e}^{-i{{\varphi }_{1}}} \\ \nonumber
 {{\Lambda }_{II-1}}&=&{{t}_{II-1kq\sigma}}{{e}^{-i\left( {{\varphi }_{I}}-{{\varphi }_{I-1}} \right)}} \\ \nonumber
 {{\Lambda }_{DN}}&=&{{t}_{DNkq\sigma}}{{e}^{i{{\varphi }_{N}}}} 
,\end{eqnarray}
and used  $\varphi_S=\varphi_D=0$.
\subsection{Integration over Grassmann fields}
\label{subsec:integration}
The Green's functions of the lead electrons are defined by
\begin{eqnarray} \label{GsubJ}
(\partial_{\tau}-\mu+\epsilon_{Jk\sigma})g_{Jk\sigma}(\tau,\tau^{\prime})=\delta(\tau-\tau^{\prime}),
\end{eqnarray}
and
\begin{eqnarray}\label{GsubJ2}
g_{Jk\sigma}(0,\tau^{\prime})= -g_{Jk\sigma}(\beta,\tau^{\prime}),
\end{eqnarray}
where $\mu$ denotes the chemical potential of the lead $J$. Correspondingly, for the island electrons we have
\begin{eqnarray}\label{GsubI}
(\partial_{\tau}-\mu+\epsilon_{Iq\sigma})G_{Iq\sigma}(\tau,\tau^{\prime})=\delta(\tau-\tau^{\prime})
\end{eqnarray}
and
\begin{eqnarray}\label{GsubI2}
G_{Iq\sigma}(0,\tau^{\prime})= -G_{Iq\sigma}(\beta,\tau^{\prime}),
\end{eqnarray}
where $\mu$ also denotes the chemical potential of the island $I$. The solution of the inhomogeneous differential equation in Eq.(\ref{GsubI}) with the boundary condition
in Eq.(\ref{GsubI2}) reads \cite{Negele1987}
\begin{eqnarray}\label{GsubI3}
G_{Iq\sigma}(\tau,\tau^{\prime})=\left\{ \begin{array}{ccc}
{\displaystyle\frac{\exp\left[-\epsilon^{}_{Iq\sigma}(\tau-\tau^{\prime})\right]}{1+\exp\left(-\beta\epsilon^{}_{Iq\sigma}\right)}} &
\hbox{for} &  \tau>\tau^{\prime}\, , \\ & &  \\
-{\displaystyle\frac{\exp\left[-\epsilon^{}_{Iq\sigma}(\tau-\tau^{\prime})\right]}{1+\exp\left(\beta\epsilon^{}_{Iq\sigma}\right)}} &
\hbox{for} & \tau<\tau^{\prime}\, ,
\end{array} \right.
\end{eqnarray}
with the analogous expression for $g_{Jk\sigma}(\tau,\tau^{\prime})$. For convenience, hereafter, we have measured all energies from the chemical potential of the leads and islands.

The action in Eq.(\ref{Zchi}) is quadratic in the Grassmann fields $\vec{\zeta}$ and $\vec{\theta}$ as can be seen explicitly form Eq.(\ref{Selectron}). Hence, the corresponding path integral is Gaussian and can be performed analytically. Thus, we can integrate over the quasi-particle reservoirs in three steps as in the following. The first step, 
using the general Gaussian's integral formula as \cite{Negele1987}
\begin{eqnarray}\label{Gaussmatrix}
    \int\prod\limits_{i=1}^{N}{\frac{d{{x}_{i}}^{*}\,d{{x}_{i}}}{2\pi i}}\,{{e}^{-x_{i}^{*}\,{{H}_{ij\,}}{{x}_{j}}+\,J_{i}^{*}{{x}_{i}}+\,x_{i}^{*}{{J}_{i}}}}\,\,\,\,={{\left( \det \left( H \right) \right)}^{-1}}{{e}^{\,J_{i}^{*}\,H_{ij}^{-1}\,{{J}_{i}}}},
\end{eqnarray}
we can integrate over the fermion in the source electrode in Eq.(\ref{Zchi}) as
\begin{eqnarray}
 \int{D\mu \left({{\zeta }_{S}} \right)}\,{{e}^{-S[\vec{\zeta},\vec{\theta},\vec{\varphi}]}}&=&\int{D\mu \left({{\zeta }_{S}} \right)}\,{{e}^{-\left( \zeta _{S}^{*}g_{S}^{-1}{{\zeta }_{S}}+\theta _{1}^{*}\Lambda _{1S}^{*}{{\zeta }_{S}}+{{\zeta }^{*}}_{S}\Lambda _{1S}\theta _{1} \right)}} \\ \nonumber
 &=& Z_{S}\,e^{\theta _{1}^{*}\Lambda _{1S}^{*}g_{S}\Lambda _{1S}\theta _{1}},
\end{eqnarray}
where ${{Z}_{S}}=\det \left( g_{S}^{-1} \right)$. The second step, we integrate over ${{\theta }_{1}}$ as
\begin{eqnarray} \label{zeta1}
  \int{D\mu \left({{\theta }_{1}} \right)}\,{{e}^{-S[\vec{\zeta},\vec{\theta},\vec{\varphi}]}}&=&\int{D\mu \left({{\theta }_{1}} \right)}\,{{e}^{-\left( \theta _{1}^{*}G_{1}^{-1}{{\theta }_{1}}+\theta _{2}^{*}\Lambda _{21}^{*}{{\theta }_{1}}+{{\theta }^{*}}_{1}\Lambda _{21}^{{}}\theta _{2}^{{}} \right)+\theta _{1}^{*}\Lambda _{1S}^{*}{{g}_{S}}\Lambda _{1S}^{{}}\theta _{1}^{{}}}} \nonumber \\
 &=&\det \left( G_{1}^{-1}-\Lambda _{1S}^{*}{{g}_{S}}\Lambda _{1S}^{{}} \right)\,{{e}^{\theta _{2}^{*}\Lambda _{21}^{*}\tilde{G}_{1}^{{}}\Lambda _{21}^{{}}\theta _{2}^{{}}}} \\ \nonumber
 &=&{{Z}_{1}}\det \left( 1-G_{1}^{{}}\Lambda _{1S}^{*}{{g}_{S}}\Lambda _{1S}^{{}} \right)\,{{e}^{\theta _{2}^{*}\Lambda _{21}^{*}\tilde{G}_{1}^{{}}\Lambda _{21}^{{}}\theta _{2}^{{}}}},
\end{eqnarray}
where ${{Z}_{1}}=\det \left( G_{1}^{-1} \right)$ and $\tilde{G}_{1}^{-1}=G_{1}^{-1}-\Lambda _{1S}^{*}{{g}_{S}}\Lambda _{1S}$.
We have used the identity matrix properties $G_1^{-1}G_1=\mathbb{I}$, and the determinant matrix property, respectively.
In the same way, one integrates over $\theta_I$ variable and then obtains
\begin{eqnarray} \label{zetaI}
 \int{D\mu({\theta }_{I}})\,e^{-S[\varphi, \Lambda]}&=&Z_{I}\det( 1-G_{I}\Lambda _{II-1}^*\tilde{G}_{I-1}\Lambda _{II-1}) \\
 &&\times e^{\theta _I^*\Lambda _{II-1}^*\tilde{G}_{I-1}\Lambda _{II-1}\theta _{I-1}}  \nonumber,
\end{eqnarray}
where ${{Z}_{I}}=\det \left( G_{I}^{-1} \right)$, and 
\begin{eqnarray} \label{GItil}
    \tilde{G}_{I}^{-1}=G_{I}^{-1}-\Lambda _{II-1}^{*}{{\tilde{G}}_{I-1}}\Lambda _{II-1}.
\end{eqnarray} 
The third step, the fermion in drain $\zeta_D$ can be integrated as
\begin{eqnarray}
     \int{D\mu(\zeta_{D})}\,e^{-S[\varphi, \Lambda]}=Z_{D}\det( 1-{g_{D}}\Lambda _{DN}^{*}{\tilde{G}}_{N}\Lambda _{DN})\,,
\end{eqnarray}
where $Z_D=\det(g_D^{-1})$ and $\tilde{G}_{N}^{-1}=G_{N}^{-1}-\Lambda _{NN-1}^{*}{{\tilde{G}}_{N-1}}\Lambda _{NN-1}$.
We may rewrite the partition function in Eq.(\ref{ZTB}) as
\begin{eqnarray}\label{Zold}
 Z_{BT}\left[ \vec{\varphi }\right]&=&\mathcal{N}_{BT}\,\det \left( 1-G_{1}\Lambda _{1S}^{*}g_{S}\Lambda _{1S} \right)\, \\ \nonumber
 &&\times \prod\limits_{I=2}^N \det \left( 1-G_{I}\Lambda _{II-1}^{*}{\tilde{G}_{I-1}}\Lambda _{II-1} \right) \\ \nonumber
 &&\times\det \left( 1-{{g}_{D}}\Lambda _{DN}^{*}{\tilde{G}_{N}}\Lambda _{DN} \right)\,,
\end{eqnarray}
where $\mathcal{N}_{BT}=Z_S Z_D Z_1 Z_2...Z_N$. To simplify future, we rewrite the partition function in Eq.(\ref{Zold}) in term of the trace of matrix as
\begin{eqnarray}
 Z_{BT}\left[ \vec{\varphi }\right]&=&\mathcal{N}_{BT}{{e}^{tr\big\{\ln \left( 1-G_{1}\Lambda _{1S}^{*}{{g}_{S}}\Lambda _{1S} \right) \big\}}}{{e}^{tr\big\{ \ln \left( 1-{{g}_{D}}\Lambda _{DN}^{*}{{\tilde{G}}}_{N}\Lambda _{DN}^{{}} \right) \big\}}} \\ \nonumber
 &&\times {{e}^{\sum\limits_{I=2}^{N}{tr\big\{ \ln \left( 1-G_{I}^{{}}\Lambda _{II-1}^{*}{{{\tilde{G}}}_{I-1}}\Lambda _{II-1} \right) \big\}}}}
\end{eqnarray}
 where we have used the matrix property as
\begin{eqnarray}
    \det A={{e}^{tr\left\{ \ln A \right\}}}.
\end{eqnarray}{}
We can therefore rewrite the partition function of the system in term of the effective action as
\begin{eqnarray} \label{Zbt}
    Z[\vec{\varphi}]=\mathcal{N}_{sys}\prod\limits_{I=1}^N\bigg[\sum_{k_{I}=-\infty}^{\infty}\int\limits_{\varphi_{I}(0)}^{\varphi_{I}(\beta)+2\pi
k_{I}}D[\varphi_{I}(\tau)]\bigg]{{e}^{-{{S}_{\text{eff}}\left[\vec{\varphi},  \Lambda \right]}}},
\end{eqnarray}{}
where $\mathcal{N}_{sys}=\mathcal{N}\mathcal{N}_{BT}$ and $S_{\text{eff}}\left[\vec\varphi,\Lambda\right]$ is the effective action of the island system
\begin{eqnarray}
    S_{\text{eff}}\left[\vec\varphi, \Lambda \right]={{S}_{C}}\left[\vec{\varphi}\right]+{{S}_{T}}\left[\vec\varphi, \Lambda\right],
\end{eqnarray}
with the Coulomb action ${{S}_{C}}\left[ \vec{\varphi}  \right]$ obtains in Eq.(\ref{Sc}) and the tunneling action is expressed as
\begin{align}\label{St}
 S_{T}\left[ \vec \varphi, \Lambda  \right]=&-tr\big\{ \ln \left( 1-G_{1}^{{}}\Lambda _{1S}^{*}{{g}_{S}}\Lambda _{1S}^{{}} \right) \big\}-\sum\limits_{I=2}^{N}{tr\big\{ \ln \left( 1-G_{I}^{{}}\Lambda _{II-1}^{*}{{{\tilde{G}}}_{I-1}}\Lambda _{II-1}^{{}} \right)\big\}} \nonumber \\
 &-tr\big\{ \ln \left( 1-{{g}_{D}}\Lambda _{DN}^{*}{{{\tilde{G}}}_{N}}\Lambda _{DN}^{{}} \right)\big\}.
\end{align} 

\subsection{The tunneling action}\label{subsec:theSt}
In order to evaluate the tunneling action in Eq.(\ref{St}) explicitly, we have followed the ideas of Ambegaokar et al.\cite{Ambegaokar1982} and Grabert \cite{Grabert1994}. We expand the first term in Eq.(\ref{St}) as
\begin{eqnarray}\label{1stTerm}
  -tr\big\{ \ln \left( 1-G_{1}^{{}}\Lambda _{1S}^{*}{{g}_{S}}\Lambda _{1S}^{{}} \right) \big\}
 &=&tr\bigg\{ \ln {{\left( 1-G_{1}^{{}}\Lambda _{1S}^{*}{{g}_{S}}\Lambda _{1S}^{{}} \right)}^{-1}} \bigg\} \\ \nonumber \\ \nonumber
 &=&tr\bigg\{ \sum\limits_{n = 1}^\infty  \frac{{{{\left( {{G_1}\Lambda _{1S}^*{g_S}{\Lambda _{1S}}} \right)}^n}}} {n} \bigg\} .
\end{eqnarray}
For the metallic oxide-layer tunnel junctions, the tunneling conductance arises from a large number, i.e., $ M=\sum_{\sigma } {1}$ of available tunneling channels while the tunneling amplitude and, accordingly, the matrix element of the tunneling matrices $(\Lambda)$ introduced in Eq.(\ref{Lamda def}) are very small. It is sufficient that one may keep only the first term in the expansion in Eq.(\ref{1stTerm}) as
\begin{eqnarray} \label{FirstTerm}
  -tr\{ \ln (1-G_{1}\Lambda _{1S}^{*}{{g}_{S}}\Lambda _{1S} )\}\approx tr\{ G_{1}\Lambda _{1S}^{*}{{g}_{S}}\Lambda _{1S} \}.
\end{eqnarray}
This approximation is known as the large channel number approximation \cite{Grabert1994}, and will be discussed in more detail later.
To evaluate the trace of the matrix in Eq.(\ref{FirstTerm}), we substitute $\Lambda_{1S}$ and $\Lambda_{1S}^*$ from (\ref{Lamda def}) into (\ref{FirstTerm}) as
\begin{equation} \label{tr1}
  tr \{ G_{1}\Lambda _{1S}^{*}{{g}_{S}}\Lambda _{1S} \}
 =\sum\limits_{kq\sigma }{\int\limits_{0}^{\beta }{d\tau }\int\limits_{0}^{\beta }{d{\tau }'}}{{\left| {{t}_{1Skq\sigma}} \right|}^{2}}{e^{i(\varphi_1(\tau)-\varphi_1(\tau'))}}{{G}_{1kq}}\left( \tau ,{\tau }' \right){{g}_{Sq\sigma}}\left( \tau ,{\tau }' \right),
\end{equation}
where ${\left| {{t_{1Skq\sigma}}} \right|^2} = t_{1Skq\sigma}^*{t_{1Skq\sigma}}$. We introduce the electron-hole-pair propagator as
\begin{eqnarray}\label{PropLS}
\Pi_{1S}(\tau,\tau^{\prime})=\sum_{kq\sigma}|t_{1Skq\sigma}|^{2}G_{1k\sigma}(\tau,\tau^{\prime}) g_{Sq\sigma}(\tau,\tau^{\prime}) .
\end{eqnarray}
By inserting the Green's function $G_{1k\sigma}(\tau, \tau')$ in Eq.(\ref{GsubI3}) and the analogous expression for $g_{Sq\sigma}(\tau, \tau')$, the propagator $\Pi_{1S}$ becomes
\begin{eqnarray}
\Pi_{1S}(\tau,\tau^{\prime})=-\sum_{kq\sigma}|t_{1Skq\sigma}|^{2}
\frac{\exp\left[-\epsilon^{}_{1k\sigma}(\tau-\tau^{\prime})\right]}{1+\exp\big(\mp
 \beta\epsilon^{}_{1k\sigma}\big)}
 \frac{\exp\left[-\epsilon^{}_{Sq\sigma}(\tau^{\prime}-\tau)\right]}{1+\exp\big(\pm
 \beta\epsilon^{}_{Sq\sigma}\big)}\, ,
\end{eqnarray}
where the upper and lower sign holds for $\tau > \tau^{\prime}$ and $\tau < \tau^{\prime}$, respectively. Here we have measured all energies from the chemical potential of the leads and islands.  Clearly,
$\Pi_{1S}(\tau,\tau^{\prime})=\Pi_{1S}(\tau-\tau^{\prime})$ depends only on $\tau-\tau^{\prime}$.

The tunnel matrix elements may be approximated as a constant in the relevant energy range, closed to the Fermi energy at low temperature. Thus, the energy dependence of the tunneling amplitudes can be neglected for energies near the Fermi energy, which contribute to the energy integrals. The propagator $ \Pi_{1S}$ can be transformed to read 
\begin{eqnarray}\label{PropLS2}
  \Pi_{1S}(\tau-\tau')=-|t_{1S}|^{2}\int_{-\infty}^{\infty}d\epsilon \int_{-\infty}^{\infty}d\epsilon^\prime \sum_{\sigma} \rho_{\sigma} \rho'_{\sigma} \frac{{\rm
  e}^{-\epsilon(\tau-\tau')}}{1+{\rm e}^{\mp \beta\epsilon}}\,    \frac{{\rm e}^{\epsilon^{\prime}(\tau-\tau')}}{1+{\rm e}^{\pm
  \beta\epsilon^{\prime}}} \, ,
\end{eqnarray}
where the summation over wave vectors were transformed as the integration over energies. We have used a constant density of state per channel, i.e., $\rho_{\sigma}\equiv\rho_{\sigma}(0)$ and extended the limit of the integration since the metallic bandwidth is much larger than the relevant energy scales, $E_C$ and $k_B T$ \cite{Grabert1994}. The integration over the energies can be performed \cite{Gradshteyn1994} and by replacing $\rho_{\sigma}$ and $\rho'_{\sigma}$ with the average densities of state per channel $\rho$ and $\rho'$ of the island and the lead, one gets
\begin{eqnarray}\label{PropLS3}
  \Pi_{1S}(\tau-\tau') = -g_{1S}\, \alpha(\tau-\tau'), 
\end{eqnarray}
where we have introduced the tunneling kernel 
\begin{eqnarray}
  \alpha \left( {\tau  - \tau '} \right) = \frac{1}{ {4{\beta}^2\sin^2\left( {\frac{\pi }{\beta }(\tau-\tau') } \right)}} ,
\end{eqnarray}
and  the dimensionless tunneling conductance
\begin{eqnarray}\label{tunnelG1S}
g^{}_{1S}= 4\pi^2|t_{1S}|^{2}M \rho \rho'
\end{eqnarray}
of the tunnel junction between the source electrode and the first island. Accordingly, the tunneling conductance $g^{}_{1S}$ can be treated as a constant and 
we insert (\ref{PropLS3}) into Eq.\,(\ref{tr1}) to yield
\begin{eqnarray} \label{GT2:6}
tr \{ G_{1}\Lambda _{1S}^{*}{{g}_{S}}\Lambda _{1S} \}=- g_{1S}\int_0^{\beta}d\tau\int_0^{\beta}d\tau'\,\alpha(\tau-\tau')\cos\left[\varphi_1(\tau)-\varphi_1(\tau')\right], 
\end{eqnarray}
where the imaginary part vanishes because of $\alpha(\tau-\tau^{\prime})$ is an even function.

Analogous to the calculation of the first term above, one can evaluate the second and third terms in Eq.(\ref{St}) with the approximation of $\tilde{G}_I$ as
\begin{eqnarray}\label{Gapprox}
    {\tilde{G}}_{I}={(1+G_{I}\Lambda _{II-1}^{*}G_{I-1}\Lambda _{II-1})}G_{I},
\end{eqnarray}
where we have approximated $\tilde{G}_{I}^{-1}$ in Eq.(\ref{GItil}) by the inverse matrix property and
\begin{eqnarray}
    \frac{1}{1-x}=\sum\limits_{n=0}^{\infty }{{{x}^{n}}}\approx(1+x).
\end{eqnarray}
Using the definition of $\Lambda_{II-1}$ in Eq.(\ref{Lamda def}) and (\ref{Gapprox}), one can obtain the propagators
\begin{eqnarray}\label{alphadnii-1}
    {\Pi_{II-1}}\left( \tau-\tau' \right)=-g_{II-1}\, \alpha(\tau-\tau'),
\end{eqnarray}
and
\begin{eqnarray}\label{alphadn}
     {\Pi_{DN}}\left(\tau-\tau' \right)=-g_{DN}\, \alpha(\tau-\tau'),
\end{eqnarray}
with
\begin{eqnarray}\label{tunnelGII-1}
g_{II-1}= 4\pi^2|t_{II-1}|^{2}M \rho \rho',
\end{eqnarray}
and
\begin{eqnarray}\label{tunnelGDN}
g_{DN}= 4\pi^2|t_{DN}|^{2}M \rho \rho' \, .
\end{eqnarray}
At this point, we will explain the large channel number approximation in more detail, clarifying why the higher order terms in the expansion of the logarithm in Eq.(\ref{St}) may be neglected in the limit of $ M \gg 1$. The form the definitions in Eq.(\ref{PropLS3}), (\ref{alphadnii-1}), and (\ref{alphadn}) shows that by keeping only the first order of the expansion, the dimensionless conductance is proportional to $ M|t_{II-1}|^2$. It is easy to see that for the higher order terms, the term $M|t_{II-1}|^{n}$ is proportional to $ g_{II-1}^{n} M^{-(n-1)} $ for any integer number $n \geq 2$, which is negligible for the large $M$ \cite{Goppert1999}.      
The dimensionless conductance can be related to the conductance of the tunnel junction between two conductors $I$ and $I-1$ as
\begin{eqnarray}
    {g}_{I I-1}=\frac{2\pi }{{{e}^{2}}}{{G}_{N N-1}},
\end{eqnarray}
with $g_{1S}$ and $g_{DN}$ are related to ${G}_{1S}$ and ${G}_{DN}$,  respectively.

By the large channel number approximation, the tunneling action in Eq.(\ref{St}) can be rewritten as 
\begin{eqnarray}\label{St2}
 S_{T}[\vec{\varphi }] &=& -\int\limits_{0}^{\beta }{d\tau }\int\limits_{0}^{\beta }{d{\tau }'}\alpha(\tau -{\tau }' )\bigg\{ g_{1S}\cos[\varphi _{1}(\tau)-\varphi_{1}({\tau }')] \\ \nonumber
 &&+ \sum\limits_{I=2}^{N}{g_{II-1}\cos[(\varphi_{I-1}(\tau)-\varphi_{I-1}({\tau }'))-(\varphi_{I}(\tau)-\varphi_{I}({\tau }'))]}  \\ \nonumber
 &&+ g_{DN}\cos[\varphi_{N}(\tau)-\varphi_{N}({\tau }')]\bigg\} ,
\end{eqnarray}
where the imaginary parts were vanished due to the integrands being odd functions. We close this section with the expressed of the partition function as
\begin{eqnarray} \label{Zsys}
    Z[\vec{\varphi}]=\mathcal{N}_{sys}\prod\limits_{I=1}^N\bigg[\sum_{k_{I}=-\infty}^{\infty}\int\limits_{\varphi_{I}(0)}^{\varphi_{I}(\beta)+2\pi
k_{I}}D[\varphi_{I}(\tau)]\bigg]{{e}^{-{{S}_{\text{eff}}\left[\vec{\varphi} \right]}}},
\end{eqnarray}{}
where the effective action reads
\begin{eqnarray} \label{Seff}
 S_{\text{eff}}[\vec{\varphi}]&=&\int\limits_{0}^{\beta }{d\tau } \bigg \{ \frac{1}{4}{{{\dot{\vec{\varphi }}^T}}}\mathbf{E}_{CN}^{-1}\dot{\vec{\varphi }}-i \vec{n}_{g}^{T}\cdot \dot{\vec{\varphi }} \bigg \} \\ \nonumber
 &&- \int\limits_{0}^{\beta }{d\tau }\int\limits_{0}^{\beta }{d{\tau }'}\alpha(\tau -{\tau}')\bigg\{g_{1S}\cos[{{\varphi }_{1}}(\tau)-{{\varphi }_{1}}({{\tau }'})]\\ \nonumber
 &&+\sum\limits_{I=2}^{N}{g_{II-1}\cos[({{\varphi }_{I-1}}(\tau)-{\varphi_{I-1}}({\tau }'))-(\varphi _{I}(\tau)-\varphi_{I}({\tau }'))]}  \\ \nonumber
 &&+{{g}_{DN}}\cos[{{\varphi }_{N}}(\tau)-{{\varphi }_{N}}( {{\tau }'})]\bigg\} ,
\end{eqnarray}
with $\mathbf{E}_{CN}^{-1}$ is related to the matrix defined in Eq.(\ref{ECN}).

\section{Applications}
\label{sec:applications}

The previous section found that the partition function corresponding to the path integral is a non-Gaussian integral and cannot be performed analytically. However, since all possible paths are expressed in imaginary time, the quantum mechanical propagator turns into a quantum statistical density matrix to perform the quantum Monte Carlo method \cite{Ceperley1995, Negele1987, Grotendorst2002, Troyer2005}. Therefore, to show an application of the partition function in Eq.(\ref{Zsys}) with the effective action in Eq.(\ref{Seff}),  we will propose the calculation of average electron numbers of the serial island system and its application in the following.

\subsection{Dimensionless Energies}
For more convenience in numerical calculations, such as the quantum Monte Carlo simulation of the single-electron transistor \cite{walli02}, one may measure all energies in the relevant charging energy scale unit. Thus, in principle, one is free to choose the reference energy scale. However, this paper will use the charging energy that may be defined as
\begin{eqnarray} \label{Ec}
    {E_C} = {G^{cl}}\sum\limits_{I=1}^{N+1} {\frac{{E_{II-1}}}{{{G_{II-1}}}}},
\end{eqnarray}
for the serial island system in Fig \ref{fig:fig_model}. ${G^{cl}}$ denotes the total of high-temperature conductance of all tunneling junctions, i.e.,  
\begin{eqnarray}
    {G^{cl}} = {\left( {\sum\limits_{I=1}^{N+1} {\frac{1}{{{G_{II-1}}}}} } \right)^{ - 1}},
\end{eqnarray}
where $G_{II-1}$ stands for the high-temperature conductance between island $I\, \text{and}\, (I-1)$ with $G_{10}=G_{1S}$ and $G_{N+1 N}=G_{DN}$. The coefficient $E_{II}$ is the element of the matrix $E_{N}$ defined in Eq.(\ref{ECN}). In the unit of the charging energy ($E_C$) given in Eq.(\ref{Ec}), one can rewrite the Coulomb action in Eq.(\ref{Sc}) as
\begin{eqnarray}\label{SC}
    S_{C} [\vec{\varphi}(\tau)]=\int_{0}^{\beta E_C} d\tau
\bigg \{ \frac{1}{4}\dot{\vec{\varphi}}^{T}\mathbb{E}_{N}\,\dot{\vec{\varphi}}+i (\vec{ n}^{T}_{g}\cdot\dot{\vec{\varphi}}) \bigg \} ,
\end{eqnarray}
where 
\begin{eqnarray}\label{Efinal}
    \mathbb{E}_{N}=\frac{2 E_C}{e^2}\mathbf{C}_N\equiv
\begin{pmatrix}
  \mathbb{E}_{11} & \mathbb{E}_{12} & \cdots & \mathbb{E}_{1N} \\
  \mathbb{E}_{21} & \mathbb{E}_{22} & \cdots & \mathbb{E}_{2N} \\
  \vdots & \vdots & \vdots & \vdots \\
  \mathbb{E}_{N1} & \cdots & \cdots & \mathbb{E}_{NN}
\end{pmatrix},
\end{eqnarray}
and the matrix $\mathbf{C}_N $ is defined in Eq.(\ref{Cn}). The tunneling action in Eq.(\ref{St2}) can be rewritten as
\begin{eqnarray} \label{st}
 S_{T}[\vec{\varphi}(\tau)] &=& -\int\limits_{0}^{\beta E_C}{d\tau}\int\limits_{0}^{\beta E_C}{d\tau'}\alpha(\tau-\tau')\bigg\{ g_{1S}\cos[\varphi _{1}(\tau)-\varphi_{1}(\tau')] \\ \nonumber
 &&+ \sum\limits_{I=2}^{N}{g_{II-1}\cos[(\varphi_{I-1}(\tau)-\varphi_{I-1}(\tau'))-(\varphi_{I}(\tau)-\varphi_{I}(\tau'))]} \\ \nonumber
 &&+ g_{DN}\cos(\varphi_{N}(\tau)-\varphi_{N}({\tau }')) \bigg\} ,
\end{eqnarray}
 with the tunneling kernel
\begin{eqnarray}
  \alpha \left( {\tau  - \tau '} \right) = \frac{1}{ {4({\beta}E_C)^2\sin^2 \left( {\frac{\pi }{\beta E_C }(\tau-\tau') } \right)}} .
\end{eqnarray}
 
\subsection{Winding Numbers}
The partition function in Eq.(\ref{Zsys}) is expressed as the sum over paths with different boundary conditions. Instead of evaluating each path integral separately up to a certain
cutoff and adding them up. In numerical calculations, it is more convenient to make the transformation
\begin{eqnarray}
    {\varphi _I}\left( \tau  \right) = {\xi _I}\left( \tau  \right) + \nu_{k_{I}}\tau,
\end{eqnarray}{}
where $\nu_{k_{I}} = {{2\pi {k_I}} \mathord{\left/
 {\vphantom {{2\pi {k_I}} {\left( {\beta {E_C}} \right)}}} \right.
 \kern-\nulldelimiterspace} {\left( {\beta {E_C}} \right)}}$ with all periodic paths obey the condition,
 \begin{eqnarray}
     {\xi _I}\left( 0 \right) = {\xi _I}\left( {\beta {E_C}} \right).
 \end{eqnarray}{}
Using this transformation, one can rewrite the partition function in Eq.(\ref{Zsys}) in suitable form as
\begin{eqnarray}\label{Zchi3}
Z[\vec{\xi},\vec{k}]&=&\mathcal{N}_{sys}\prod\limits_{I=1}^N\bigg[\sum_{{k_I}=-\infty}^{\infty}\int\limits_{\xi_{I}(0)}^{\xi_{I}(\beta {E_C})}D[\xi_{I}(\tau)]\bigg]{\rm e}^{-S_{\rm eff}[\vec{\xi},\vec{k}] },
\end{eqnarray}
where the effective action can be expressed in term of the winding numbers as
\begin{eqnarray}\label{seff}
    S_{\rm eff}[\vec{\xi},\vec{k}]=S_C[\vec{\xi},\vec{k}]+S_T[\vec{\xi},\vec{k}].
\end{eqnarray}
The Coulomb action in Eq.(\ref{SC}) can be rewritten as
\begin{eqnarray}\label{SC2}
S_{C} [\vec{\xi},\vec{k}]=\int_{\xi(0)}^{\xi({\beta E_C})} d\tau\,\frac{1}{4}\vec{\dot{\xi}}^{T}\mathbb{E}_{N}\,\vec{\dot{\xi}}+\frac{4\pi^2}{\beta E_C}\vec{k}^T \mathbb{E}_{N}\,\vec{k}+2\pi i(\vec{ n}^{T}_{g}\cdot\vec{k}), 
\end{eqnarray}
where  $\vec{\xi}(\tau)=(\xi_1(\tau),...,\xi_N(\tau))^{T}$  with given winding numbers, $\vec{k}=(k_1,...,k_N)^{T}$. The tunneling action in Eq.(\ref{st}) can be rewritten as 

\begin{eqnarray}\label{stxi}
S_{T}[ {\vec \xi},\vec{k}] &=&  - \int\limits_{\xi(0)}^{\xi(\beta {E_C})} {d\tau } \int\limits_{\xi(0)}^{\xi(\beta {E_C})} {d\tau '} \alpha \left( {\tau  - \tau '} \right)\bigg\{{g_{1S}}\cos \left[{{\xi _1}\left( \tau  \right) - {\xi _1}\left( {\tau '} \right) + {\nu _{k_1}}{k_1}} \right] \nonumber \\
   &&+ \sum \limits_{I=2}^N {g_{II - 1}}\cos \left[({\xi_{I-1}}(\tau)-{\xi_{I-1}}(\tau'))-({\xi_{I}}({\tau} )-{\xi_{I}}({\tau'})) \right. \hfill \\ \nonumber
   &&\,\,\,\,\,\,\,\,\,\,\,\,\,\,\,\,\,\,\,\,\,\,\,\,\,\,\,\,\,\,\,\,\,\,\,\,\,\,\,\,\,+ (\nu _{k_{I - 1}}- \nu _{k_I})(\tau - \tau')] \\ \nonumber
   &&+ {g_{DN}}\cos \left[{{\xi _{N}}\left( \tau  \right) - {\xi _{N}}\left( {\tau '} \right) + {\nu _{k_N}}{k_N}} \right] \hfill \bigg \}. \nonumber
\end{eqnarray}
\subsection{Average electron number}
In this section, we apply the partition and the effective action in Eqs.(\ref{Zchi3})--(\ref{stxi}) to calculate average electron numbers in the serial island system. The total electron number of the serial island system may be defined as
\begin{eqnarray}\label{ntotal}
  \left\langle {{n}_{Total}} \right\rangle =\sum\limits_{I=1}^{N}{\left\langle {{n}_{I}} \right\rangle },
\end{eqnarray}
where $\left\langle {{n}_{T}} \right\rangle$ denotes the total average electron number of the system and $\left\langle {{n}_{I}} \right\rangle$ denotes the average electron number of the island $I$. Analogous to the definition of charge fluctuations in the single electron box \cite{Grabert1994, Goppert2001}, the average electron number on island $I$ can be expressed as 
\begin{eqnarray}\label{nisland}
    \langle {{n_I}} \rangle  = {n_{0I}} + \frac{{1}}{{2\beta E_C}}\Big(\frac{\partial \ln Z }{\partial n_{0I }}\Big). 
\end{eqnarray}
By inserting the partition function in Eq.(\ref{Zchi3}) into (\ref {nisland}), we can express the average electron number on islands into three cases, i.e.,
\begin{eqnarray} \label{n1} 
    \left\langle {{n}_{1}} \right\rangle={n_{01}} - \frac{{4\pi}i}{{\beta E_C}}\left( {{\mathbb{E}_{11}}\left\langle {{{k}_{1}}} \right\rangle + {\mathbb{E}_{12}}\left\langle {{{k}_2}} \right\rangle } \right),
\end{eqnarray}

\begin{eqnarray} \label{ni}
    \left\langle {{n}_{I}} \right\rangle={n_{0I}} - \frac{{4\pi}i}{{\beta E_C}} \left({\mathbb{E}_{I I }}\left\langle {{{k}_I}} \right\rangle + {{\mathbb{E}_{II - 1}}\left\langle {{{k}_{I-1}}} \right\rangle + {\mathbb{E}_{II+1}}\left\langle {{{k}_{I + 1}}} \right\rangle } \right),
\end{eqnarray}
and
\begin{eqnarray} \label{nN}
    \left\langle {{n}_{N}} \right\rangle={n_{0N}} - \frac{{4\pi}i}{{\beta E_C}}\left( {{\mathbb{E}_{NN}}\left\langle {k_N} \right\rangle + {\mathbb{E}_{NN-1}}\left\langle {{{k}_{N-1}}} \right\rangle } \right),
\end{eqnarray}
for the first, intermediate, and last island, respectively. The coefficients in Eqs.(\ref{n1})-(\ref{nN}) are elements of the matrix obtained in Eq.(\ref{Efinal}). The expectation value of the winding numbers $\left\langle k \right\rangle$ in Eqs.(\ref{n1})-(\ref{nN}) is defined as
\begin{eqnarray}\label{kI}
    \left\langle k_{I} \right\rangle  = \frac{\prod\limits_{I=1}^N\bigg[\sum_{{k_I}=-\infty}^{\infty}\int\limits_{\xi_{I}(0)}^{\xi_{I}(\beta E_C)}D[\xi_{I}(\tau)]\bigg]{ {{k_{I} \, {e^{ - {S_{{\rm{eff}}}}\left[ \vec \xi ,\vec k \right]}}} } }}{\prod\limits_{I=1}^N\bigg[\sum_{{k_I}=-\infty}^{\infty}\int\limits_{\xi_{I}(0)}^{\xi_{I}(\beta E_C)}D[\xi_{I}(\tau)]\bigg]{ {{{e^{ - {S_{{\rm{eff}}}}\left[ \vec \xi ,\vec k \right]}}} } }},
\end{eqnarray}
where the effective action is obtained in Eqs.(\ref{seff})-(\ref{stxi}). In analytical calculation, one cannot obtain an exact solution of the expectation value of the winding number in Eq.(\ref{kI}) because it is a non-Gaussian integral. However, by quantum Monte Carlo calculation, one can evaluate the expectation value of the winding number and the average electron number of the island system, as an example in the following section. Moreover, in the Monte Carlo calculation of the expectation value in Eq.(\ref{kI}), the fermionic sign problem \cite{Ceperley1995} would arise from the imaginary part of the Coulomb action in Eq.(\ref{SC2}). To reduce the influence of the fermionic sign problem, one can apply the idea in Ref. \cite{Troyer2005} to perform the quantum Monte Carlo simulation.  
\subsection{Two-island system}
\begin{figure}
\centering
\includegraphics[width=0.5\textwidth]{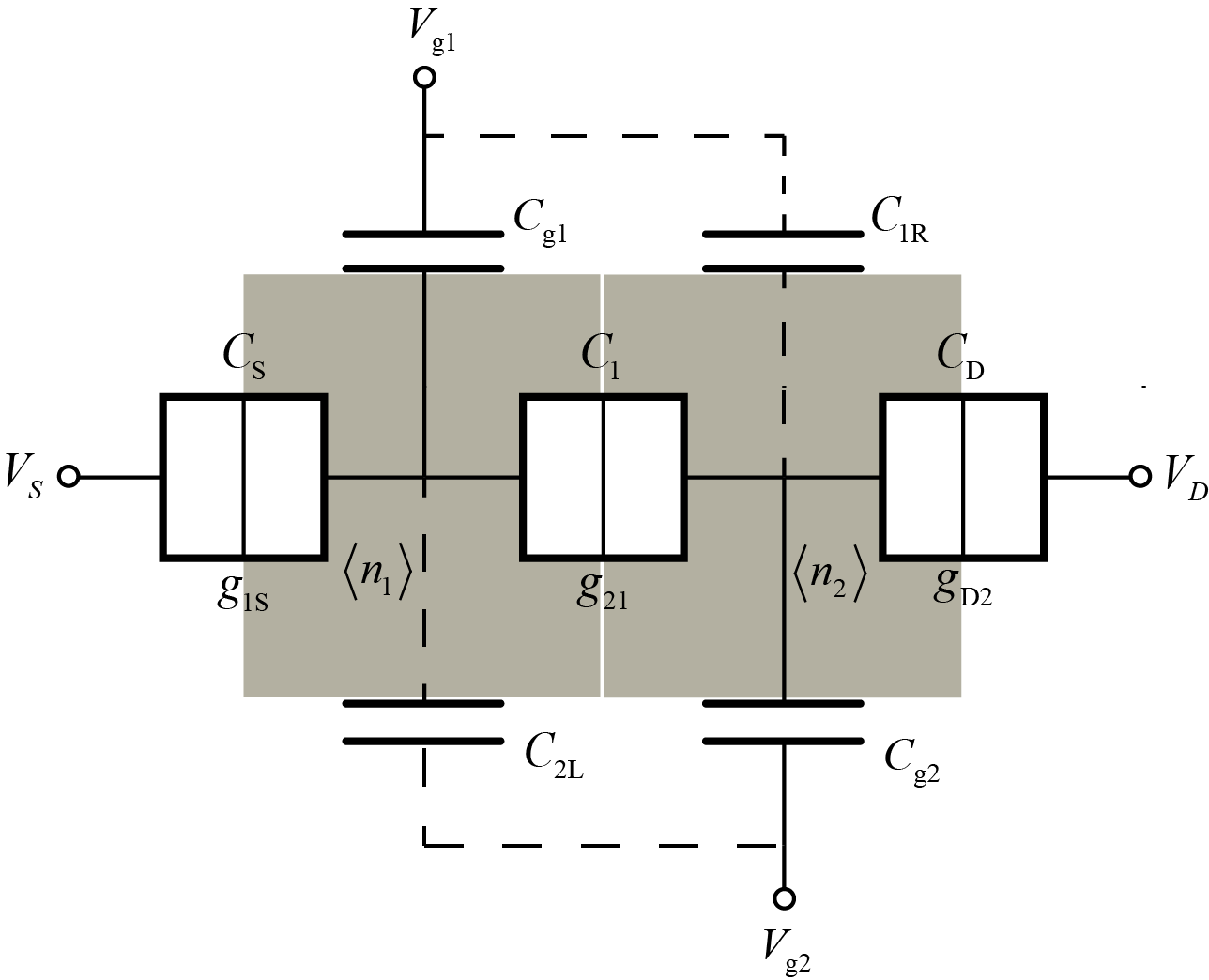}
\vspace{0.2cm}
\caption{ The equivalent circuit diagram of the two-island system \cite{limbach05}.} 
\label{fig:SEP}
\end{figure}

In this section, we focus on the two-metallic island system \cite{limbach05,limbach02} and calculate average electron numbers of the system by the Monte Carlo method. Limbach \textit{et al.} \cite{limbach05} reported the experimental results of two-metallic island systems, usually called the single electron-pump (the SEP), which can be represented by the circuit diagram in Fig.\,\ref{fig:SEP}. This arrangement is biased by the voltage difference $V_{D}-V_{S}$ denoted by $V_{DS}$. Two gate voltages can tune the electrostatic potentials on two islands, i.e., $V_{g1}$ and $V_{g2}$, which couple directly to the islands by the capacitance $C_{g1}$ and $C_{g2}$. The experimentally unavoidable stray capacitors are represented by $C_{2L}$ and $C_{1R}$. We conclude this experiment with Table \ref{ex:parameters} containing all parameters that one needs to simulate the average electron number of the SEP.
\label{subsec:esetup}

\begin{table}[!h] \label{para2island}
\begin{center}
\begin{tabular}{|c||p{0.7cm}|p{0.7cm}|p{0.7cm}|p{0.7cm}|p{0.7cm}|p{0.7cm}|p{0.7cm}|p{0.7cm}|p{0.7cm}|p{0.7cm}|p{0.7cm}|}
\hline
 Parameters& $C_{S}$ & $C_{1}$ & $C_{D}$& $C_{g1}$ & $C_{1R}$ & $C_{2L}$ & $C_{g2}$ & $g_{1S}$ & $g_{21}$ & $g_{D2} $ & $G^{cl}_{2} $   \\
 \hline \hline
 Values& 181 & 173 & 236 & 50.5  & 18.0 & 21.5  & 58.6 & 0.52 & 1.32 & 0.83 & 10.0  \\
 \hline
 Units& (aF) & (aF) & (aF) & (aF)  & (aF) & (aF)  & (aF) & - & - & - & $(\mu S)$  \\
 \hline
 \end{tabular}
 \end{center}
 \caption{\footnotesize Parameters of the SEP\,\cite{limbach02}. A dimensionless conductance of the individual tunneling junction is defined as $g_{j}=G_{j}/G_{K}$ where $j \in \{1S,21,D2\}$, $G_{K} = e^2/h$, and $ G^{cl}_{2}$ stands for the high-temperature conductance of the SEP. The charging energy $E_C$ defined as in Eq.(\ref{Ec}) is equal to $0.184\, meV$.}
\label{ex:parameters}
\end{table}

From the definition of the total average electron number in Eq.(\ref{ntotal}), we can rewrite the total average electron number for the two-island system as
\begin{eqnarray} \label{nT2island}
    \left\langle {{n_{Total}}} \right\rangle  = \left\langle {{n_1}} \right\rangle  + \left\langle {{n_2}} \right\rangle,
\end{eqnarray}
where the average electron number of the first and second island can be obtained by
\begin{eqnarray}\label{n12island}
    \langle {{n_1}} \rangle  = {n_{01}} - \frac{{4\pi}i}{{\beta E_C}}( {{\mathbb{E}_{11}} \langle {{{k}_1}} \rangle + {\mathbb{E}_{12 }}\langle {{{k}_2}} \rangle } ),
\end{eqnarray}
and
\begin{eqnarray}\label{n22island}
    \langle {{n_2}} \rangle  = {n_{02}} - \frac{{4\pi}i}{{\beta} E_C}( {{\mathbb{E}_{22}}\langle {{{k}_2}} \rangle + {\mathbb{E}_{21}}\langle {{{k}_1}}\rangle } ),
\end{eqnarray}
where the induced charges on the first and second islands are generally denoted by $n_{01}=(C_{S}V_{S}+C_{1}V_{1}+C_{g1}V_{g1}+C_{2L}V_{g2})/e$, and $n_{02}=(C_{1}V_{1}+C_{D}V_{D}+C_{g2}V_{g2}+C_{1R}V_{g1})/e$, respectively. The expectation value of the winding number $\left\langle {{k_I}} \right\rangle$ of the two-island system is expressed as
\begin{eqnarray}\label{kI2}
    \left\langle {{k_I}} \right\rangle  = \frac{{ {\sum_{{k_1}, {k_2}}
\int\limits_{\xi_{1}(0)}^{\xi_{1}(\beta E_C)}\!\!\!\!
D[\xi_{1}(\tau)]\!\!\int\limits_{\xi_{2}(0)}^{\xi_{2}(\beta E_C)}\!\!\!\! D[\xi_{2}(\tau)]{k_I}{e^{ - {S_{{\rm{eff}}}}\left[ \vec \xi ,\vec k \right]}} } }}{{ {\sum_{{k_1}, {k_2}}
\int\limits_{\xi_{1}(0)}^{\xi_{1}(\beta E_C)}\!\!\!\!
D[\xi_{1}(\tau)]\!\!\int\limits_{\xi_{2}(0)}^{\xi_{2}(\beta E_C)}\!\!\!\! D[\xi_{2}(\tau)]{e^{ - {S_{{\rm{eff}}}}\left[ \vec \xi ,\vec k \right]}} } }},
\end{eqnarray}
where the effective action reads
\begin{eqnarray}
    S_{{\rm{eff}}}\left[ \vec \xi ,\vec k \right]= S_{C} [\vec{\xi},\vec{k}]+ S_{T}[ {\vec \xi},\vec{k}],
\end{eqnarray}
and the Coulomb action 
\begin{eqnarray}\label{SCn2}
    S_{C} [\vec{\xi},\vec{k}]=\int_{\xi(0)}^{\xi({\beta E_C})} d\tau\,\frac{1}{4}\vec{\dot{\xi}}^{T}\mathbb{E}_{2}\,\vec{\dot{\xi}}+\frac{4\pi^2}{\beta E_C}\vec{k}^T \mathbb{E}_{2}\,\vec{k}+2\pi i(\vec{ n}^{T}_{g}\cdot\vec{k}),
\end{eqnarray}
where $\vec{\xi}=(\xi_1, \xi_2)^T$,  $\vec{k}=(k_1, k_2)^T$, and $\vec{ n}^{T}_{g}=(n_{01}, n_{02})$.
For the two-island system, the matrix in Eq.(\ref{Efinal}) is reduced to be
\begin{eqnarray}
    {\mathbb{E}_{2}} = \frac{2E_{C}}{e^2}\mathbf{C}_2\equiv
    \begin{pmatrix}
    {{\mathbb{E}_{11}}}&{{\mathbb{E}_{12}}} \\
    {{\mathbb{E}_{21}}}&{{\mathbb{E}_{22}}} \\
    \end{pmatrix},
\end{eqnarray}
with the capacitance matrix 
\begin{eqnarray} \label{C2}
 \mathbf{C}_{2}
=
\begin{pmatrix}
  C_{\sum 1} & -C_{1} \\
  -C_{1}     & C_{\sum 2},
\end{pmatrix},
\end{eqnarray} 
where $C_{\sum 1}=C_S+C_1+C_{g1}+C_{2L}$ and $C_{\sum 1}=C_1+C_D+C_{g2}+C_{1R}$. The tunneling action of the two-island system reads 
\begin{align}\label{STn2}
   S_{T}[ {\vec \xi},\vec{k}] =&  - \int\limits_{\xi(0)}^{\xi(\beta {E_C})} {d\tau } \int\limits_{\xi(0)}^{\xi(\beta {E_C})} {d\tau'} \alpha ({\tau  - \tau'})\bigg\{{g_{1S}}\cos [{{\xi_1}(\tau) - {\xi_1}({\tau'}) + {\nu_{k_{1}}}{k_1}}] \nonumber \\
   &+ {g_{21}}\cos[({\xi_{1}}(\tau)-{\xi_1}(\tau'))-({\xi_{2}}({\tau} )-{\xi_{2}}({\tau'})) + (\nu_{ k_{1}}- \nu_{k_{2}})(\tau - \tau')] \nonumber \\ 
   &+ {g_{D2}}\cos[{{\xi _{2}}(\tau) - {\xi _{2}}({\tau'}) + {\nu_{ k_{2}}}{k_2}}]\bigg \}.
\end{align}
\begin{figure}
\centering
\includegraphics[width=0.95\textwidth]{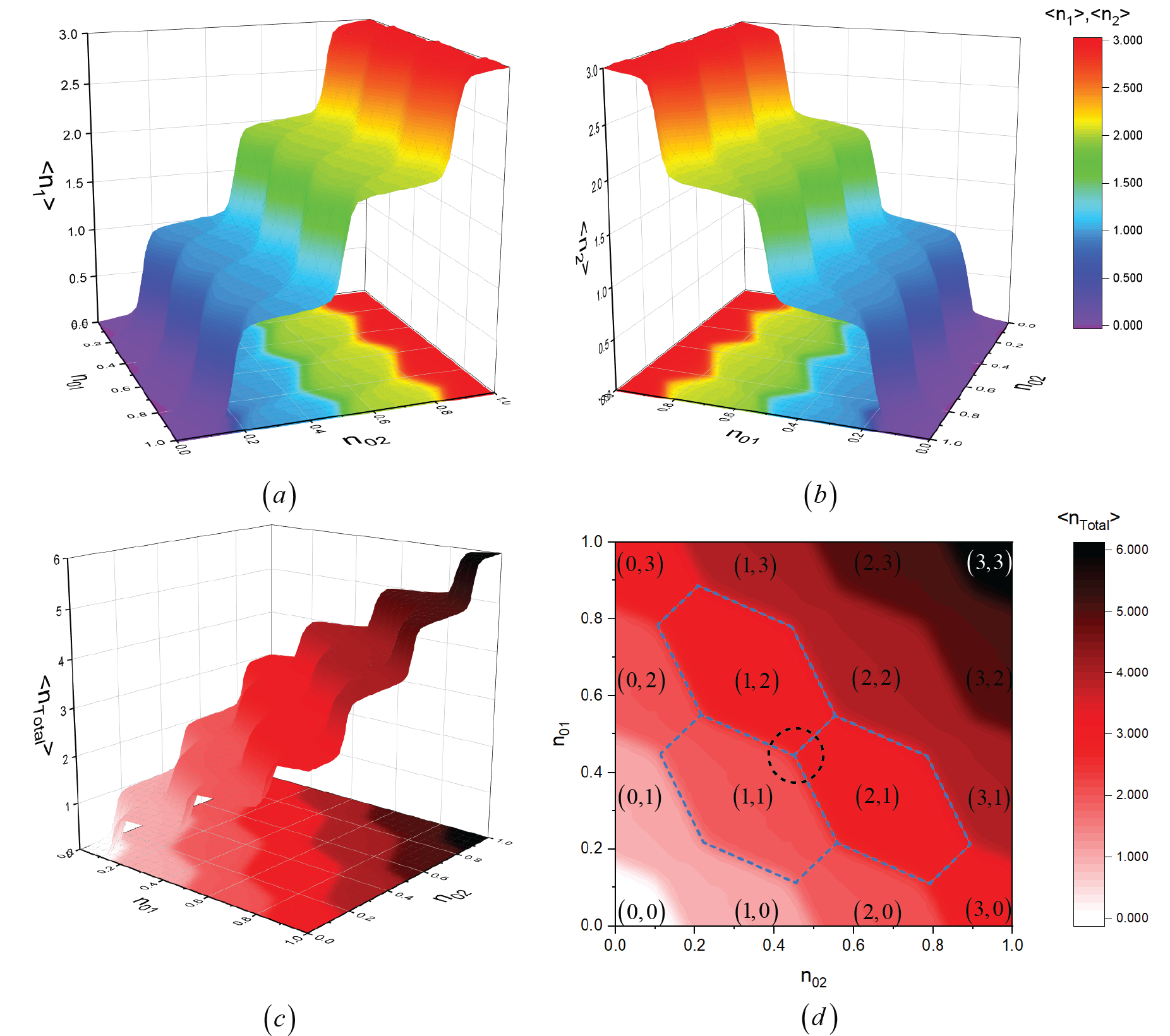}
\vspace{0.2cm}
\caption{ The average electron numbers of the first, second, and two islands are shown in Figures \ref{fig:cons SD full MC}a, \ref{fig:cons SD full MC}b, and \ref{fig:cons SD full MC}c,  respectively, were calculated by the quantum Monte Carlo method for $\beta E_C=20.0$, corresponding temperature $T = 0.1\,K$. The projection of the total average electron number on the dimensionless gate voltage plane ($n_{01}, n_{02}$) shows that the hexagonal domains mark regions indicated by the electron state $(n_1, n_2)$.}
\label{fig:cons SD full MC}
\end{figure}

Using the parameters in Table \ref{ex:parameters}, we have calculated the average electron numbers in Eqs.(\ref{nT2island})-(\ref{n22island}) by the quantum Monte Carlo method.  Fig.\ref{fig:cons SD full MC}a and \ref{fig:cons SD full MC}b show that the average electron number on the first and second islands are step functions of two dimensionless gate voltage variables. For the condition $ V_{DS} =0$, when two gate voltages are increased, the total average electron rises. This situation is well known as the Coulomb staircase and the essential behavior of a single-electron box \cite{Grabert1994, Goppert2001}. The total average electron number of the system is also the Coulomb staircase shown in Fig.\ref{fig:cons SD full MC}c. In this case, therefore, the single-electron pump behaves like a single-electron box consisting of two coupling islands. Then, electrons can propagate between the two islands, controlling with the two gates.
 
Furthermore, we found that the projection of the total average electron number on the dimensionless gate voltage plane shows hexagonal domains as in 
Fig.\ref{fig:cons SD full MC}d. The average electron numbers on each island take a fixed value indicated by the numbers in the hexagonal domains. Let us consider the triple point in the black circle, shown in Fig.\ref{fig:cons SD full MC}d, where three adjacent states can be occupied. Circling the triple point counter-clockwise corresponds to the sequence $(1,1)\rightarrow (2,1) \rightarrow(1,2)$, which describes an electron that can transfer from left to right. Therefore, in the presence of $V_{DS}$, current can flow through the two islands from the source lead to the drain lead by applying the gate voltages corresponding to the position of the triple point in the $(n_{01},n_{02})$ plane. According to the experimental results, the maximum conductance peaks of the SEP occurred near the triple points \cite{limbach05}.

To verify the hexagonal domains, we have calculated the charge stability diagram of the two-island system by the method in Ref.\cite{Fujisawa2002}, which is called the traditional stability diagram. This stability diagram was calculated by concerning only the charging energy of the system, in which the degeneracy lines between the stable charge regions depend on two dimensionless gate voltages parameters, as shown by the doted lines in Fig. \ref{fig:cons SD full MC}d. We emphasize that the tunneling effect and temperature dependence were neglected in calculating the traditional stability diagram \cite{Fujisawa2002}. As a result, we found that the doted lines overlap the borders of the hexagonal domains for low temperatures. In other words, it showed that we could construct the stability diagram of the two-island system using the average electron numbers, including the tunneling effect, which is then called the quantum stability diagram. Since single-electron transport in the island systems includes the tunneling process, the quantum stability diagram could become a powerful tool to study the island systems beyond the classical picture. 
\section{Conclusions}
\label{sec:conclusion}
In this paper, we have calculated the grand canonical partition function of the serial metallic island system by the imaginary-time path integral formalism. By the large channel approximation, the partition function as a path integral over phase fields with a path probability given the effective action functional. Furthermore, we have proposed calculating average electron numbers of the metallic island system and rewritten them in suitable forms for the quantum Monte Carlo simulation. For the demonstration, we have calculated the average electron numbers for the SEP. The results show that the average electron numbers increase with the two gate voltage variables as a step function. Therefore, the single-electron pump behaved like the single-electron box consisting of two coupling islands, wherein electrons can propagate between the two islands, controlling by the two gates. In addition, we have proposed the method to construct the stability diagram for a finite temperature, including the tunneling effect, by a projection image of the total average electron number on the dimensionless gate voltage variables plane. As a result, the stability diagram could describe the occurrence of the Coulomb blockade regions in agreement with the traditional stability diagram, in the limit of low temperature. Finally, we anticipate that the partition function will be helpful as a starting point for a further theoretical investigation into the serial island system. It would also be interesting to use the approaches described in this paper to describe single-electron device's experiments.

\section*{Acknowledgments}
\label{sec:acknow}
We would like to acknowledge financial support from the
Theoretical Condensed Matter Physics Research Unit(TCMPRU),
Maha-Sarakham University, and Thailand Research Fund
(Grant no.TRG5680021). Especially, we thank A.R. Plant for critical reading of the manuscript.

\section*{References}

\bibliographystyle{iopart-num}
\bibliography{Reference}

\end{document}